\documentstyle[preprint,aps,eqsecnum,epsf,floats,prd]{revtex}

\def\slash#1{#1\hskip-0.45em /}
\def\dslash{\partial\hskip-0.6em /}
\def\CA{{\cal A}}

\def\msc{m_\varphi}
\def\leftrightnabla{\tensor{\bnabla }}
\def\OMIT#1{{}}

\def\scP{\varphi_P}
\def\scR{\varphi_R}

\def\psieft{\psi_h}
\def\chieft{\chi_h}

\def\pvec{{\bf p}}
\def\bsigma{\mbox{\boldmath $\sigma$}}
\def\bnabla{\mbox{\boldmath $\nabla $}}

\begin{document}
\tighten
\preprint{\vbox{
\hbox{UTPT--97-12}
\hbox{DOE/ER/41014-21-N97}
\hbox{hep-ph/9707313} }}

\title{Power Counting in Dimensionally Regularized NRQCD}

\author{Michael Luke$^a$ and  Martin J. Savage$^b$ \bigskip}
\address{(a) Department of Physics, University of Toronto, Toronto, Ontario,
Canada M5S 1A7
\\  (b) Department of Physics, University of Washington, Seattle, WA 98195-1560
U.S.A.
\bigskip}

\date{July 1997}

\maketitle

\begin{abstract}
\tighten
We present a scheme for calculating in NRQCD with consistent
power counting in the heavy quark velocity $v$.  As an example,
we perform the systematic matching of an external current onto NRQCD
at subleading order in $v$, a calculation
relevant for the process $e^+e^-\rightarrow\mbox{hadrons}$ near threshold.
Consistent velocity power counting in
dimensional regularization is achieved by including two distinct gluon
fields, one corresponding to gluon radiation and one corresponding to an
instantaneous
potential.   In this scheme power counting is manifest in any gauge, and
also holds for non-gauge
interactions. The matching conditions for an external vector current in
NRQCD are
calculated to $O(g^2v^2)$ and the cancellation of infrared divergences in the
matching conditions is shown to require both gluon
fields.  Some subtleties arising in the
matching conditions at subleading order are addressed.
\end{abstract}
%\draft
\pacs{13.20.He, 12.38.Bx, 13.20.Fc, 13.30.Ce}

\section{Introduction}

Non-relativistic QCD (NRQCD) \cite{nrqcd} is a powerful tool for analyzing the
dynamics of
systems with two or more heavy quarks at momentum transfers much less
than  their
mass.   Such systems are more complicated than single-heavy quark systems
described by the
heavy quark effective theory (HQET) \cite{hqet} because the quarks scatter via
the QCD potential.
This introduces infrared divergences in heavy quark scattering near
threshold in HQET
which must be regulated by
resumming an infinite number of insertions of the kinetic energy operator.
Since
this operator is subleading in $1/m_Q$, this violates HQET power counting.
This kinematic regime of QCD is of interest for a number
of physical systems, including  quarkonium, $e^+ e^-\rightarrow\mbox{hadrons}$
near threshold and nonrelativistic QCD sum rules \cite{nrsum}.  Similar
techniques
are also of interest for other nonrelativistic systems, such as positronium
\cite{posit} and low-energy nucleon-nucleon scattering \cite{nnscatt}.

\begin{figure}[bh]
\epsfxsize=8 cm
\hfil\epsfbox{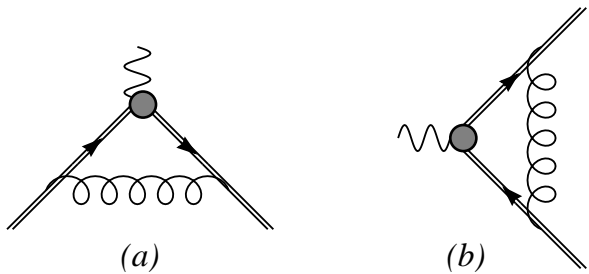}\hfill
\caption{Two of the one loop contributions to the matching of an external
current in
HQET.}
\label{hqetloop}
\end{figure}

A concrete example where HQET power counting fails
is provided by an external current in QCD coupled to $\bar q\Gamma q$,
where $\Gamma$ is some Dirac matrix.
In processes with a single incoming and outgoing heavy quark (such as
semileptonic $b\rightarrow c$ decay) there is no
potential scattering and HQET is the appropriate low-energy theory.  Loop
graphs such as Fig.\ \ref{hqetloop}(a) are well-defined in HQET, and the
matching conditions for this current in HQET are currently known to
$O(\alpha_s,1/m_Q^2)$ \cite{hqetmatch}.  On the other hand,  the
one-loop correction to quark-antiquark production by the same current
near threshold cannot be correctly described in HQET;  the one-loop graph in
Fig.\ \ref{hqetloop}(b) is infinite when the four-velocities of the quarks
are the same, and gives rise to the well-known infinite complex anomalous
dimension for the current
\cite{hqetcomplex}.
The appropriate low-energy theory for the second process is NRQCD, which treats
potential scattering near threshold properly.

In this paper we consider the general problem of matching an external
fermion-antifermion production current  in a nonrelativistic theory.   We pay
particular attention to power counting in the NRQCD expansion parameter
$v$, the relative three-velocity of the heavy quarks.
Power counting in NRQCD is less transparent than in HQET, for several
reasons.  First, since $v$ is not a parameter in the Lagrangian, power
counting is not manifest in the NRQCD Lagrangian, although there are
simple rules for determining
the $v$ scaling of an operator \cite{le92}.  Second,
the power counting for on-shell gluons
differs from that of virtual gluons contributing to
potential scattering \cite{la96,lm96,gr97} and in order to have simple $v$
counting this distinction must be implemented at the level of the Lagrangian.
Also, in order to retain simple $v$ counting beyond tree level the theory must
be
regulated with a mass-independent regulator such as dimensional regularization,
instead of the usual momentum cutoff (otherwise divergent loop integrals
change the power counting of Feynman graphs in the effective theory).

The paper is organized as follows.  In Section \ref{vpc} we discuss
velocity power counting in NRQCD, and the relation between the results
of Refs.\ \cite{lm96} and \cite{gr97}.   In order to maintain manifest $v$
power counting in
dimensionally regulated NRQCD we introduce two sets of gluon fields,
corresponding to propagating gluons and an instantaneous potential.  In
Section \ref{nrym} we discuss the matching of  an external current
in nonrelativistic Yukawa theory (NRY), and show that the dependence on
the infrared regulator vanishes in the matching when both gluon fields are
included in the low-energy theory.  We consider this theory both because it is
simpler than QCD, as well as to stress that manifest velocity power counting
does
not depend on working in
any particular gauge, such as Coulomb gauge.  In Section \ref{nrqcdm} we
match an external current in NRQCD
to $O(g^2 v^2)$, and explicitly show that the low-energy theory reproduces the
nonanalytic behaviour of QCD to this order.   Finally, in Section \ref{conc} we
present our conclusions.

\section{Velocity Power Counting}\label{vpc}

In NRQCD, the power counting of terms in the Lagrangian is different from HQET,
allowing
potential scattering near threshold to be correctly described.    Operators are
classified
according to how they scale with the three-velocity $v$ instead of $1/m_Q$
\cite{le92}.  Since the
kinetic energy of a nonrelativistic state is proportional to $v^2$
while the momentum is proportional to $v$, space and time derivatives scale
differently with $v$, and
power counting is not manifest in the NRQCD Lagrangian,
\begin{eqnarray}
{\cal L} & = &
\psieft^\dagger\left(i D_0 + {1\over 2m_Q}{\bf D}^2\right)\psieft  - {1\over 4}
G^{\mu\nu a}
G_{\mu\nu a} +{\cal L}_{\rm g.f.} + \dots\ .
\end{eqnarray}
Nonrelativistic fields are distinguished here from fields in the full theory by
the subscript $h$.
${\cal L}_{\rm g.f.}$ is the gauge fixing term, $D_\mu =
\partial_\mu+ i g_s A_\mu$ is the gauge-covariant derivative ($A_\mu\equiv
A_\mu^a T^a$)
and the dots denote higher dimension operators whose matrix elements are
suppressed by powers of $v$.

In Ref. \cite{lm96} a rescaling of the fields and  coordinates was introduced
to make the
$v$ counting of operators manifest at the level of the Lagrangian.    The NRQCD
Lagrangian
was written in terms of new coordinates ${\bf X}$ and $T$ and new fields
$\Psi_h$ and ${\cal
A}$, where
\begin{equation}
\label{lmrescale}
{\bf x} = \lambda_x {\bf X},\ \ t=\lambda_t T,\ \ \psieft =
\lambda_Q \Psi_h,
\ A^0 = \lambda_{A^0} {{\cal A}}^0,\  A^i = \lambda_{{\bf  A}} {{\cal A}}^i\,,
\end{equation}
and where $\lambda_x=1/m_Q v$, $\lambda_t=1/m_Q v^2$,
$\lambda_Q=\lambda_x^{-3/2}$ and
$\lambda_A = \lambda_{A^0} = \left( m_Q \lambda_x^3 \right)^{-1/2}$. In this
form  the $v$
scaling of operators is manifest in the Lagrangian.  In Lorentz gauge,
\begin{eqnarray}\label{rescaled1}
{{\cal L}}^R &=&\Psi_h^\dagger
\left(i\partial_0 - {g\over
\sqrt{ v }} {{\cal A}}_0  \right) \Psi_h \nonumber- {1\over2} \Psi_h^\dagger
\left(i\bnabla
- g
\sqrt{v} {\bf{{\cal A}}} \right)^2 \Psi_h \nonumber \\ && - {1\over 4} \left(
\partial_i
{{\cal A}}_j^a -
\partial_j {{\cal A}}_i^a -  g \sqrt{v} f_{abc} {{\cal A}}_i^b {{\cal A}}_j^c
\right)^2 + {1\over 2} \left(\partial_i {{\cal A}}_0^a - v  \partial_0 {{\cal
A}}_i^a - g
\sqrt{v} f_{abc} {{\cal A}}_i^b {{\cal A}}_0^c \right)^2 \nonumber \\
&&-{1\over
2\alpha}\left(v\partial^0 {{\cal A}}_0^a+\partial^i{{\cal A}}_i^a\right)^2
\nonumber \\
&=&  \Psi_h^\dagger \left(i\partial_0 + {\bnabla ^2\over 2}-{g\over \sqrt{ v }}
{{\cal A}}_0 \right)\Psi_h  - {1\over 4} \left( \partial_i {{\cal A}}_j^a -
\partial_j {{\cal A}}_i^a\right)^2+  {1\over 2} \left(\partial_i {{\cal
A}}_0^a\right)^2\nonumber \\
&&-{1\over 2\alpha}\left(\partial^i{{\cal
A}}_i^a\right)^2+O(v, g\sqrt{v}).
\end{eqnarray}

Of course, there is no physics in a simple rescaling.  However, in order to
have an effective theory in which the $v$ power counting is manifest, the
additional prescription that terms which are subleading in $v$ be treated as
operator
insertions must be added (otherwise, loop graphs evaluated in dimensional
regularization will mix powers
of $v$).  As was noted in \cite{lm96}, once this prescription is added, the
rescaling in
Eq.~(\ref{rescaled1})
misses important physics.
While it provides the correct description of virtual gluon
exchange corresponding to an  instantaneous potential, it fails to correctly
describe
on-shell gluons. The problem is that the pole in the full gluon
propagator occurs at
$k_0^2={\bf k}^2$, or, in terms of rescaled variables, $v^2 K_0^2={\bf K}^2$.
Unless the
time derivative in the gluon kinetic term is  treated exactly instead of as an
insertion
(which would violate manifest $v$ counting), amplitudes in the effective theory
do not have
the correct branch cut corresponding to physical gluon propagation, so  the
effective theory
cannot describe on-shell gluons.

In Ref. \cite{gr97} it was demonstrated that this problem is avoided by a
further
rescaling
(in Coulomb gauge) of the
space coordinates of only the transverse components of the gauge fields.  In
the language of Ref. \cite{lm96}\footnote{The
Lagrangian in \cite{gr97} was written as an expansion in powers of $1/c$
instead of $v$;
however, the two descriptions are equivalent.  Note that as
$c\rightarrow\infty$, $\alpha_s=g^2/4\pi c\rightarrow 0$, whereas as
$v\rightarrow 0$,
$\alpha_s$ remains constant, so factors of $\alpha_s$ scale differently
with $1/c$ than with $v$.}, this corresponds to the rescaling
\begin{equation} {\bf  x}=\lambda_t {\bf Y},\ t=\lambda_t T,\ A^i= \lambda_t
^{-1}\tilde
\CA^i.
\end{equation}
The kinetic term for the $\tilde\CA_i$'s is canonically normalized,
\begin{equation} L_{kin}= -\int d^3 Y\, dT\ \left[  {1\over
4}(\partial_i\tilde\CA_j^a-\partial_j\tilde\CA_i^a- g f_{abc} \tilde
\CA_i^b\tilde
\CA_j^c)^2+{1\over 2}\left(\partial_0 \tilde\CA_j^a\right)^2\right]\,,
\end{equation}
while the transverse gluon-quark interaction may be expanded in terms of
multipoles\footnote{At the level of Feynman diagrams, the observation that
on-shell
gauge fields couple via the multipole expansion  was made in Ref.\
\cite{la96}.}
\begin{eqnarray}\label{multipole} {\cal L}_{int}& = &-{i\over
2}gv\Psi_h^\dagger({\bf
X},T)\leftrightnabla_i
\Psi_h({\bf X},T)\tilde\CA_i(v{\bf X},T)
\nonumber \\ & \ =&-{i\over 2}gv\Psi_h^\dagger({\bf X},T)\leftrightnabla_i
\Psi_h({\bf X},T)\left[1+v X\cdot\bnabla +\dots\right]\tilde\CA_i(0,T).
\end{eqnarray}
Note that three-momentum is not conserved at the multipole interaction vertex,
since the
theory breaks translational invariance, although energy is conserved.
Once the multipole expansion is performed, loop integrals in dimensional
regularization do not change the $v$ scaling of a graph determined by the
vertices.

There is, however, a subtlety arising from the multipole expansion.
Since three-momentum is not conserved at the vertices,
transverse gluons cannot alter the three-momenta of
the heavy quarks.  Potential scattering via transverse gluon exchange
therefore does not occur in the low energy theory.   Since the amplitude for
potential scattering is not analytic in the external momenta,
it cannot be reproduced  in NRQCD by the addition of local operators.
Both potential scattering and real gluon emission are long-distance effects, so
in order
to correctly describe the infrared physics of QCD in the nonrelativistic limit
the
instantaneous potential due to transverse gluon exchange must be added to the
effective theory.  This may either be done by explicitly including spatially
non-local operators in the nonrelativistic theory (the need for which was
discussed
in \cite{gr97}), or by reintroducing
a second gluon field which couples according to Eq.\ (\ref{rescaled1}).

The need for two distinct gluon fields is understood  by comparing the
energy and
momentum of radiated gluons
compared to those involved in potential scattering.  In the nonrelativistic
regime of QCD
there are nonanalytic contributions to scattering amplitudes arising from
gluons in two separate kinematic regions, both with energy of order $mv^2$.
The gluons with spatial momenta $\sim mv$ are far off shell and contribute to
potential scattering.   In contrast, the gluons with spatial momenta
$\sim mv^2$ may be on-shell, describing real radiation, but do not contribute
to the scattering of quarks with three-momenta of order $mv$  (in the limit
$v\rightarrow 0$).  Each of the rescalings discussed above treats one of these
kinematic regimes
correctly, while missing the physics of the other.    As argued in Ref.
\cite{lm96},  it is not possible to describe both regimes via any single
rescaling.  In order to
correctly describe the infrared physics of QCD in the nonrelativistic limit two
separate gluon fields must be included, which will be referred to in this work
as
``potential" and ``radiation" gluons,
corresponding to the two different kinematic regimes described
above.

In this approach the heavy quark Lagrangian in NRQCD in Lorentz gauge is
therefore, in standard (unrescaled) units,
\begin{eqnarray}
{\cal L} &=& \psieft^{\dagger} \left( i\partial_0  + {\bnabla ^2\over
2m}-g A^0_P-g A^0_R(0,t)\right)\psieft-{1\over 4}\left(\bnabla^i {\bf
A}^j_P-\bnabla^j {\bf A}^i_P
\right)^2+{1\over 2}
\left(\bnabla A^0_P\right)^2\nonumber\\&&
-{1\over 4}G^{\mu\nu}_R G_{\mu\nu R}-{1\over 2\alpha}\left(\bnabla\cdot{\bf
A}_P\right)^2
-{1\over 2\alpha}\left(\partial_\mu A_R^\mu\right)^2+O(v, g\sqrt{v})\,,
\end{eqnarray}
where the subscripts $P$ and $R$ denote potential and radiation gluons,
$G^{\mu\nu}_R$ is the field strength tensor for radiation gluons, and $\alpha$
is
the gauge parameter.
For practical calculations, this version of the Lagrangian is much
more convenient than the rescaled version.  The rescaled theory simply
guarantees that loop graphs computed in the unrescaled theory, with the
appropriate terms treated as operator insertions, will have $v$ scaling
determined by the vertices.

\section{Yukawa Theory}\label{nrym}

Coulomb gauge is usually used in NRQCD because in this gauge $A_0$
exchange corresponds to an instantaneous potential.  Once $v$ counting is
performed
as in the previous section, potential gluon exchange is
instantaneous in all gauges for both transverse and longitudinal gluons.
More generally, potential scattering proceeds via an instantaneous interaction
even in theories with no gauge freedom.
To illustrate this in a theory which is simpler than QCD, in this section we
consider a nonrelativistic Yukawa theory (NRY) of
a massive
fermi field $\psi$ coupled to a massless scalar $\varphi$.

As a warmup for our calculation of the matching conditions for
$e^+e^-\rightarrow\mbox{hadrons}$ in QCD, we consider here
the matching conditions for an external current coupling
to $\bar\psi\gamma^\mu\psi$ in NRY.
In processes with a single incoming and outgoing fermion there is no
potential scattering and the analog of HQET for a Yukawa interaction is the
appropriate
effective theory.  However, as has been discussed, the $1/v$ behaviour of
potential scattering
near threshold cannot be correctly described in HQET.   In NRY,
the $1/v$ behaviour is reproduced by potential scalar exchange.  There are also
infrared divergences in the full theory due to soft scalar bremsstrahlung,
which are reproduced in
NRY by radiation scalars.
In this section this is demonstrated explicitly.
The theory is regulated in the ultraviolet by working in $d=4-\epsilon$
dimensions, and in the infrared by introducing a small scalar mass $\msc$.
(The theory
could be regulated in the infrared with dimensional regularization, as at
the end of this section, but this obscures the distinction between the
infrared and
ultraviolet divergences.)

The Lagrangian in the full theory is
\begin{equation}
{\cal L}=\bar\psi(i\dslash-m)\psi +
{1\over 2}\partial^\mu\varphi\partial_\mu\varphi
-g\bar\psi\psi\varphi\,,
\end{equation}
while the nonrelativistic Yukawa theory (NRY),
to leading order in the three-velocity $v$, is
\begin{equation}
{\cal L}_{\rm NR}=\psieft^\dagger\left(i\partial_0+{\bnabla ^2\over
2m}\right)\psieft
+{1\over 2}\partial^\mu\scR\partial_\mu\scR
-{1\over 2}(\bnabla \scP)^2-
g_1\psieft^\dagger \psieft\scP-
g_2\psieft^\dagger\psieft\scR(0,t)\,,
\end{equation}
where $\varphi_P$ and $\varphi_R$ are the potential and radiation scalars,
respectively.
There is a similar kinetic term for the anti-fermion field $\chieft$. (Note
that $\psieft$
annihilates incoming particles, while $\chieft$ creates outgoing
antiparticles).  At tree
level $g_1=g_2=g$.

The external vector current in the full theory,
\begin{equation}
J_{\mu}\bar\psi \gamma^\mu \psi\,,
\end{equation}
matches onto a number of terms in the low energy theory,\footnote{Spinors in
the full theory are normalized such that $\bar u_r(\pvec) u_s(\pvec)=
-\bar v_r(\pvec) v_s(\pvec)=\delta_{rs}$.  The two-component spinors
are $u_{h_1}=v_{h_2}=\left(1\atop0\right)$,
$u_{h_2}=v_{h_1}=\left(0\atop1\right)$.}
\begin{equation}
J_i\left(\chieft^\dagger\bsigma^i
\psieft+\psieft^\dagger\bsigma^i\chieft\right)+J_0\left(\psieft^\dagger\psieft+
\chieft^\dagger\chieft\right)+O(g^2, v^2).
\end{equation}

\begin{figure}[tb]
\epsfxsize=12 cm
\hfil\epsfbox{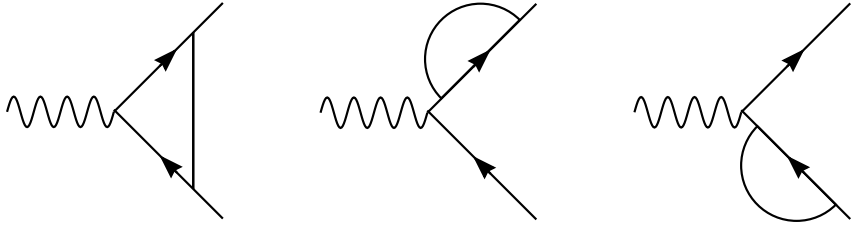}\hfill
\caption{One loop diagrams in Yukawa theory.}
\label{yukawafull}
\end{figure}

The one-loop matrix element of the current in the full theory is given by the
diagrams in Fig.\ \ref{yukawafull}.
Performing the loop integration, the infrared divergent part of the vertex
graph is
\begin{equation}
i\CA^V=4 i g^2 m^2 J_i\,\bar u(p_1) \gamma^i v(p_2)
\int{d^d k\over (2\pi)^d}{1\over (k^2-\msc^2)(k^2+2 p_1\cdot
k)(k^2-2 p_2\cdot k)}+\dots\,,
\end{equation}
where  terms finite as $\msc\rightarrow 0$ have been neglected.
Combining denominators in the usual way and performing the loop and Feynman
parameter
integrals, the infrared divergent term is found to be
\begin{equation}\label{irterm}
i\CA^V=
{g^2 \over 4\pi^2} J_i\, \bar u\gamma^i
v{(1-\beta^2)\over\beta}\tanh^{-1}\left({1\over\beta}\right)
\ln \msc  + \dots\,,
\end{equation}
where
\begin{equation}
\beta=\sqrt{1-{4m^2\over s}}
\end{equation}
is the magnitude of the three-velocity of each fermion, and $s=(p_1+p_2)^2$ is
the invariant mass of the fermion-antifermion pair.
The infrared divergent piece of the wavefunction renormalization is
\begin{equation}
i\CA^W ={g^2\over 4\pi^2}\ J_i\,\bar u\gamma^i v\ \ln \msc
+ \dots
\end{equation}
and therefore the infrared divergence in the full theory amplitude is
\begin{eqnarray}\label{fullir}
i\CA_{IR}&=&
{g^2\over 4\pi^2}J_i\,\bar u\gamma^i v\left[{1-\beta^2\over
\beta}\tanh^{-1}\left(1\over\beta\right)
+1\right]\ln \msc + \dots \nonumber \\
& = &\left[-{i g^2\over 8\pi\beta}+ {g^2\over 2\pi^2}+ O(\beta)
\right]
J_i\,\bar u\gamma^i v\,\ln \msc+\dots \nonumber \\
& = &\left[-{i g^2\over 8\pi\beta}+ {g^2\over 2\pi^2}+ O(\beta)
\right]
J_i\,u_h^\dagger\bsigma^i v_h\,\ln \msc+\dots\,,
\end{eqnarray}
where $u_h$ and $v_h$ are two-component spinors.
Note that Eq.\ (\ref{fullir}) may be written
\begin{equation}
i\CA_{IR}
={g^2\over 4\pi^2} J_i\,\bar u\gamma^i v\left[  r(w)-1  \right]
\ln\msc+\dots\,,
\end{equation}
where
\begin{equation}
w ={1+\beta^2\over -1+\beta^2},
\end{equation}
and the function $r(w)$ is given by
\begin{equation}
r(w)={1\over \sqrt{w^2-1}}\ln\left[w+\sqrt{w^2-1}\right].
\end{equation}
This is the analytic continuation to the production region
(negative $w=v\cdot v^\prime<0$) of the infrared
divergence encountered in HQET (for scalar exchange)\cite{sj92}.
The first term in Eq. (\ref{fullir}) is singular as $\beta\rightarrow 0$,
corresponding to the infinite complex
anomalous dimension found in HQET at threshold.
Since it is imaginary, it does not contribute to the decay rate at $O(g^2)$.
The second term in Eq. (\ref{fullir})
cancels in physical matrix elements with scalar bremsstrahlung.

\begin{figure}[tb]
\epsfxsize=14cm
\hfil\epsfbox{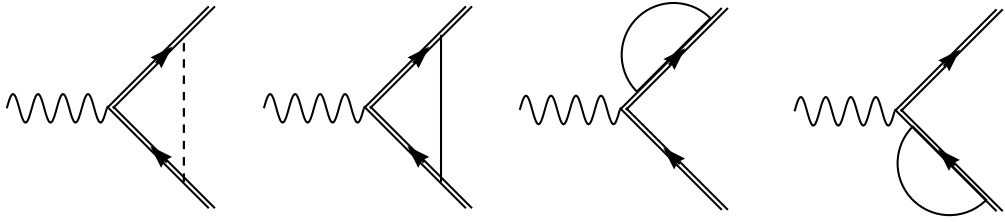}\hfill
\caption{Infrared divergent one loop diagrams in NRY.  The dashed line
corresponds to the potential scalar while the solid line is the radiation
scalar.}
\label{yukawaeft}
\end{figure}

In order to be able to match onto NRY at one loop,
both of these divergences must be reproduced in the low energy theory (the
singularities higher
order in $\beta$ will only be reproduced when operators higher order in the
velocity
are included in the effective theory).    Diagrams with both potential $\scP$
and radiation  $\scR$ scalars contribute to the amplitude in NRY (Fig.\
\ref{yukawaeft}).
The wavefunction graphs with $\varphi_P$ exchange vanish,
while the vertex graph gives
\begin{equation}
i{\cal A}^V_P=ig^2 J_i u_h^\dagger\bsigma^i v_h
\int {d^d k\over (2\pi)^d}
{1\over\left(T+k_0-{({\bf k}+{\pvec})^2\over 2m}
+i\varepsilon\right)\left(T-k_0-{({\bf k}+{\bf p})^2\over
2m}+i\varepsilon\right)
\left({\bf k}^2+\msc^2-i\varepsilon\right)}\,,
\end{equation}
where $T=E-m$ is the fermion kinetic energy.
Closing the $k_0$ integral in the upper half plane, using the leading order
equation of motion $T=p^2/2m$ and picking out
the corresponding pole leaves a $d-1$ dimensional Euclidean integral,
which gives
\begin{eqnarray}
i{\cal A}^V_P&=&g^2 m J_i u_h^\dagger\bsigma^i v_h
 {\Gamma\left(2-{d-1\over 2}\right)\over(4\pi)^{d-1}}
\int_0^1 dx \,(\msc^2(1-x)-x^2{\bf p}^2-i \varepsilon)^{{d-1\over 2}-2}
\nonumber \\
& =&-{i g^2\over 8\pi\beta} J_i u_h^\dagger\bsigma^i v_h\, \ln
\msc+\dots
\end{eqnarray}
and reproduces the first term in Eq.~(\ref{fullir}).
The radiation scalar vertex correction is
\begin{equation}
i{\cal A}^V_{R}=-ig^2 J_i u_h^\dagger\bsigma^i v_h
\int {d^d k\over (2\pi)^d}{1\over (k^2-\msc^2+i\varepsilon)
k_0^2}\,,
\end{equation}
where the $+i\varepsilon$'s in the fermion propagators have been dropped
because these
poles do not contribute to physical matrix elements.
One can see by working with
off-shell states that the poles from the fermion propagators
only give contributions
proportional to powers of $E-{\bf p}^2/2m$ which
vanish by the lowest order equations of motion of NRY.
Using the standard HQET trick of combining denominators with a
dimensionful parameter, the vertex graph becomes
\begin{eqnarray}
i{\cal A}^V_{R}& = &
8 ig^2 J_i u_h^\dagger\bsigma^i v_h
\int_0^\infty\lambda\,d\lambda\int {d^d k\over (2\pi)^d}{1\over
(-k^2-2\lambda k_0+\msc^2)^3}\nonumber \\
&=&{g^2\over 4\pi^2} J_i u_h^\dagger\bsigma^i v_h\,  \ln\msc
+\dots\ .
\end{eqnarray}
The wave function graphs give an identical contribution, so the sum of
radiation scalar graphs becomes
\begin{equation}
i{\cal A}_{R}={g^2\over 2\pi^2}J_i u_h^\dagger\bsigma^i v_h\,\ln\msc + \dots\ ,
\end{equation}
reproducing the second term in Eq.~(\ref{fullir}).

This illustrates that both $\varphi_P$ and $\varphi_R$ are required for the
difference between
the matrix elements of the external current in the full and effective theories
to be infrared finite.
Having demonstrated this, it is easier to calculate the matching conditions by
regulating both
the infrared and ultraviolet divergences in the full and effective theories
with dimensional regularization.
The matrix element in the full theory is found to be
\begin{eqnarray}\label{yukdr}
i{\cal A}_{\rm full}&=&J_i u_h^\dagger\bsigma^i v_h\left(1-{g^2\over 4\pi^2}
\left[{2\over d-4}+\gamma_E+\log{m^2\over 4\pi\mu^2}\right]\right.\nonumber\\
&&\qquad+\left.i{g^2\over 16 \pi\beta}
\left[{2\over d-4}-i\pi+\gamma_E
+\log{m^2\beta^2\over\pi\mu^2}\right]\right)+O(v).
\end{eqnarray}
Regulating the theory in both the ultraviolet and infrared with dimensional
regularization has the advantage that one-loop graphs in the nonrelativistic
theory containing
radiation scalars vanish.  This is due to a cancellation of infrared and
ultraviolet divergences, since
one-loop integrals containing radiation scalars are of the form
\begin{equation}\label{radloop}
\int {d^dk\over (2\pi)^d} f(k_0, k^2)\,,
\end{equation}
which has no mass scale and so vanishes in dimensional regularization.
Thus, radiation scalars do not contribute to the one-loop matching
conditions.  This does not mean, however, that radiation
scalars are irrelevant.  Integrals of the form
(\ref{radloop}) are both ultraviolet and infrared divergent, with the divergent
terms having
the same magnitude but opposite signs.  In the difference between the full and
effective theories the infrared divergences in the two theories cancel, leaving
an
ultraviolet divergence in the effective theory.  Unlike the infrared
divergence, the
ultraviolet divergence is cancelled in the low-energy theory by a local
counterterm.

The matrix element of the current in the effective theory with the tree-level
matching is,
in dimensional regularization,
\begin{equation}\label{nrydr}
i{\cal A}_{\rm NRY}=J_i u_h^\dagger \bsigma^i v_h\left(1+i{g^2\over 16
\pi\beta}
\left[{2\over d-4}-i\pi+\gamma_E
+\log{m^2\beta^2\over\pi\mu^2}\right]\right)+O(v).
\end{equation}
The difference between the two matrix elements (\ref{yukdr}) and (\ref{nrydr})
is analytic in the external momenta, as it must
be to be absorbed into the coefficients of local operators
in NRY.  The matching condition for the current at one loop at a
renormalization scale $\mu$ is therefore
\begin{eqnarray}
J_\mu\bar\psi\gamma^\mu\psi&\rightarrow& \left(1-{g^2\over 4\pi^2}
\left[{2\over d-4}+\gamma_E+\log{m^2\over 4\pi\mu^2}\right]\right)
\left[J_i\psieft^\dagger\bsigma^i\chieft\right]_0+\dots\nonumber\\
&&=\left(1-{g^2\over 4\pi^2}\ln{m^2\over\mu^2}
\right)J_i\psieft^\dagger\bsigma^i\chieft\,,
\end{eqnarray}
where we have denoted the bare operator by the subscript $0$, and the
effective theory
is renormalized using $\overline{\mbox{MS}}$.

%Hurrah!

\section{NRQCD}\label{nrqcdm}

The matching of an external vector current in NRQCD, relevant for
$e^+e^-\rightarrow \mbox{hadrons}$ near threshold, proceeds in much the same
way as in
the Yukawa theory of the previous section.  Infrared divergences odd in $v$ are
reproduced
in the nonrelativistic theory by potential gluon exchange, while infrared
divergences even
in $v$ are reproduced by radiation gluon exchange.  In this section the
matching conditions for a vector current in NRQCD to order $v^2$ are
calculated.\footnote{See also Ref.\ \cite{ah97},
where the matching of an external electromagnetic current to
nonrelativistic quantum
mechanics (regulated in position space) was discussed.}
The theory is regulated in both the infrared and ultraviolet in dimensional
regularization.  Cancellation in the matching conditions of terms which are not
analytic in the
external momenta is rather nontrivial at subleading order and provides a nice
demonstration of
the consistency of this approach.

\OMIT{While $S$ matrix elements are independent of gauge, the coefficients
of individual
operators in the effective theory are not necessarily gauge independent.
Coulomb gauge is particularly simple because
the radiation $A^0_R$ fields cannot be on-shell in the nonrelativistic
theory, whereas
these may in Feynman gauge.
Gauge invariance of NRQCD ensures gauge invariance
of matrix elements in NRQCD order by order in $v$, although this is no longer
manifest in the Lagrangian.}

When working at subleading orders in $v$, there are a few subtleties which must
be taken
into account.  First of all, since the three-momentum in the effective theory
is
\begin{equation}
|{\bf p}|  =  m \gamma\beta=m {\beta\over\sqrt{1-\beta^2}}\,,
\end{equation}
derivatives acting on operators in the nonrelativistic theory give factors of
$\gamma\beta$,
rather than $\beta$.  It is therefore more convenient to treat $\gamma\beta$
as the nonrelativistic expansion parameter. In the rest of this paper terms of
order $|\pvec|^n/m^n=\gamma^n\beta^n$ will be referred to as being of order
$v^n$.

Secondly,
Feynman diagrams in the full theory yield $S$ matrix elements evaluated between
relativistically normalized states, satisfying $\langle k^\prime |k\rangle=2
E_k\,(2\pi)^3
\delta^{(3)}(\vec  k-\vec k^\prime)$.  However, in the nonrelativistic theory
defined such
that the residue of the pole in the propagator is $i$, Feynman diagrams
instead give $S$ matrix elements between nonrelativistically normalized states,
defined such
that $\langle \vec k^\prime  |\vec k\rangle=2 m\,(2\pi)^3 \delta^{(3)}(\vec
k-\vec
k^\prime)$. To demonstrate this, consider the two-point functions in the full
and effective
theories.  Expanding the relativistic propagator for quarks in terms
of the
low-energy variables  gives
\begin{eqnarray}\label{twopointfull}
{i(\slash{p}+m)\over p^2 - m^2} & = &  {2i\,m\over (m+T)^2 - {\bf p}^2
- m^2} \nonumber\\ & = & {2i\,m\over 2 m T + T^2 - {\bf p}^2}\nonumber\\ & = &
{i\left(1-{{\bf p}^2\over 2m^2}\right)\over T - {{\bf p}^2\over 2m}} +{i\over T
- {{\bf
p}^2\over 2m}}\left[{i{\bf p}^4\over 8m^3}\right]
 {i\over T - {{\bf p}^2\over 2m}}
 +O(\pvec^6)\,,
\end{eqnarray}
where an irrelevant constant
term has been dropped.   This two-point function is  reproduced in the
nonrelativistic
theory by a Lagrangian
\begin{equation}\label{eftkin}
{\cal L}^\prime = \psieft^{\prime\dagger} \left(
i\partial_0
+ {\bnabla ^2\over 2m}\right)\psieft^\prime-
\psieft^{\prime\dagger} {\bnabla ^2\over 2m^2}
\left( i\partial_0 + {\bnabla ^2\over 2m}\right)\psieft^\prime +
\psieft^{\prime\dagger} {\bnabla ^4\over 8 m^3}\psieft^\prime
\end{equation}
where the field operator $\psieft^\prime$
($\psieft^{\prime\dagger}$)  annihilates (creates) a nonrelativistic particle.
However, the
residue  of the pole in
$T-{\bf p}^2/2m$ in this theory is not $i$, but $i(1-{\bf p}^2/2m+\dots)=im/E$.
While this is perfectly consistent, it is preferable to remove this extra
factor of
$m/E$.   This may be easily done, since the operator
\begin{equation}\label{singlepower}
-\psieft^{\prime\dagger} {\bnabla ^2\over 2m^2}
\left( i\partial_0 + {\bnabla ^2\over 2m}\right)\psieft^\prime
\end{equation}
is proportional to the equations of motion, and may therefore be removed by
the field redefinition
\begin{equation}
\psieft^\prime\rightarrow\psieft=\left(1-{\bnabla ^2\over
4m^2}\right)\psieft^\prime=\sqrt{E\over m}\,\psieft^\prime+O(v^4).
\end{equation}
However, because of this rescaling, an additional Feynman rule of $\sqrt{E/m}$
for each
external leg must be included in NRQCD, when evaluating matrix elements between
relativistically
normalized states.  If this term is omitted, Feynman diagrams in NRQCD
correspond to matrix elements between nonrelativistically normalized states. (A
more careful analysis using the LSZ
reduction formula in the nonrelativistic theory instead of rescaling arguments
reproduces this
result.) In the rest of the discussion the $\psieft$  fields will be used, and
matching conditions are calculated using nonrelativistically normalized states
in the full and
effective theories (this is the origin of the factors of $\sqrt{m/E}$ in the
matching
conditions presented in \cite{am97}.)

The kinetic term for the nonrelativistic fields therefore takes the usual form
\begin{eqnarray}
{\cal L}_h &=& \psieft^{\dagger} \left( i\partial_0  + {\bnabla ^2\over
2m}\right)\psieft+\chieft^{\dagger} \left( i\partial_0  - {\bnabla ^2\over
2m}\right)\chieft\nonumber\\
&&+{1\over 8m^3}\left( \psieft^\dagger \bnabla ^4 \psieft - \chieft^\dagger
\bnabla ^4 \chieft
\right)+O(v^4).
\end{eqnarray}
The low energy theory contains both potential
($A_P^\mu\equiv A_P^{\mu a}T^a$) and radiation ($A_R^\mu\equiv A_R^{\mu a}T^a$)
gluons.  The kinetic term for the gauge fields is, including the gauge
fixing terms,
\begin{eqnarray}
{\cal L}_g&=&-{1\over 4}\left(\bnabla^i {\bf A}^j_P-\bnabla^j {\bf A}^i_P
\right)^2+{1\over 2}
\left(\bnabla A^0_P\right)^2
-{1\over 2\alpha}\left(\bnabla\cdot{\bf A}_P\right)^2\nonumber\\
&&-{1\over 4}G^{\mu\nu}_R G_{\mu\nu R}
-{1\over 2\alpha}\left(\partial_i A_R^i\right)^2
+O(g\sqrt{v})\,,
\end{eqnarray}
where Coulomb gauge corresponds to the limit $\alpha\rightarrow 0$.  The
$O(g\sqrt{v})$ terms correspond to the triple-potential gluon vertex and will
not
be required for the one-loop matching we consider in this section.   Also note
that
in Coulomb gauge there is no $O(v^2)$ correction to the $A_P^0$ propagator.  In
Lorentz gauge,
the gauge-fixing term for the radiation fields is instead
$-{1\over 2\alpha}\left(\partial_\mu A_R^\mu\right)^2$, and there are
additional terms bilinear in the $A_P$'s suppressed by powers of $v$ coming
from
the expansion of the gauge-fixing term.

Working in the centre of mass frame ${\bf p}_1=-{\bf p}_2\equiv {\bf p}$,
and using the  relation (easily verified with on-shell bispinors)
\begin{eqnarray}\label{gammamatch}
\bar u(p_1)\gamma^i v(p_2)&=&u_h^\dagger\left(1+{\pvec^2\over
2m^2}\right)\bsigma^ih-{1\over
2m^2}h^\dagger \pvec\cdot\bsigma \pvec^i v_h+O(v^4)\nonumber\\
&=&{E\over
m}u_h^\dagger\left(\bsigma^i-{1\over 2m^2}\pvec\cdot\bsigma
\pvec^i\right)v_h+O(v^4)\,,
\end{eqnarray}
the tree-level matching conditions for an external vector current $J(x)$ are
found to be
\begin{eqnarray}\label{vcurrent}
J_\mu\bar\psi\gamma^\mu\psi&\rightarrow&
c_1 {\bf O}_1+c_2{\bf O}_2+c_3 {\bf O}_3+\dots+O(v^4)\nonumber\\
c_1&=&c_2=1+O(g^2)\nonumber\\ c_3&=&O(g^2)\,,
\end{eqnarray}
where
\begin{eqnarray}\label{eftops}
{\bf O}_1&=&J_i\psieft^\dagger\bsigma^i\chieft\nonumber\\
{\bf O}_2&=&{1\over 4m^2}J_i\left[\psieft^\dagger\left(\roarrow\bnabla
\cdot\bsigma\roarrow\bnabla ^i+
\loarrow\bnabla \cdot\bsigma\loarrow\bnabla ^i\right)\chieft\right]\nonumber\\
{\bf O}_3&=&{1\over 2m^2}J_i\left[\psieft^\dagger\bsigma^i\left(\roarrow\bnabla
^2+\loarrow\bnabla ^2
\right)\chieft\right]
\end{eqnarray}
and only the terms contributing to quark-antiquark production have been
included.   The operators in Eq.\ (\ref{eftops}) are renormalized in
$\overline{\mbox{MS}}$.

The $A^0_P$ coupling to heavy fermions, giving the Coulomb potential,
is $O(v^{-1/2})$,
\begin{equation}\label{lcoul}
{\cal L}_C=-g\left(\psieft^\dagger A^0_P
\psieft+\chieft^\dagger A^0_P
\chieft\right) +O(g^3).
\end{equation}
The leading corrections to this are the Darwin and spin-orbit couplings,
arising
at $O(v^{3/2})$,
\begin{eqnarray}\label{dsoterms}
{\cal L}_{D,SO}&=&\frac{g}{8m^2}
\left( \psieft^\dagger T^a \psieft+ \chieft^\dagger T^a  \chieft
\right)\bnabla ^2A^{0a}_P \nonumber\\
&&+ i\frac{g}{4m^2}\epsilon^{ijk}
\left( \psieft^\dagger T^a \bsigma^i \bnabla ^j \psieft
	+ \chieft^\dagger T^a \bsigma^i \bnabla ^j\chieft \right)\bnabla ^k
A^{0a}_P
+O(g^3).
\end{eqnarray}
Transverse potential gluons couple through the $\pvec\cdot {\bf A}$
and Fermi (chromomagnetic dipole moment) terms at $O(v^{1/2})$,
\begin{eqnarray}\label{pdotfermi}
{\cal L}_{p\cdot A,F}&=&{g\over 2m}\left(\psieft^\dagger\left({\bf A}_P\cdot
\bnabla +\bnabla \cdot {\bf A}_P\right)\psieft-\chieft^\dagger\left({\bf A}_P
\cdot\bnabla +\bnabla \cdot{\bf A}_P\right)\chieft\right)\nonumber\\
&&- \frac{g}{2m}
\left( \psieft^\dagger \bsigma \cdot (\bnabla\times {\bf A}_P)\psieft
	-\chieft^\dagger \bsigma \cdot(\bnabla\times{\bf A}_P)\chieft
\right)+O(g^3)
\end{eqnarray}
and so transverse gluon exchange and the leading relativistic corrections
to Coulomb exchange both contribute to potential scattering at $O(g^2v)$,
as expected.
This is also
the same order as the correction to
Coulomb scattering due to the $\bnabla ^4$ correction to the fermion legs.
This agrees with  the  power counting of Ref. \cite{le92} in which the matrix
element of each
of these terms is  given as $O(v^2)$, since in quarkonium  $v\sim g^2$.  Note
also that
the Fermi, Darwin, spin-orbit and relativistic kinematic corrections in the
previous equations
are only the leading pieces of the usual
form of these terms \cite{nrqcd}
\begin{eqnarray}
\delta{\cal L}_{\rm bilinear}
&=& \frac{1}{8m^3}
\left( \psi_h^\dagger ({\bf D}^2)^2 \psi_h \;-\; \chi_h^\dagger ({\bf D}^2)^2
\chi_h
\right)  \nonumber \\
&+& \frac{1}{8m^2}
\left( \psi_h^\dagger ({\bf D} \cdot g {\bf E} - g {\bf E} \cdot {\bf D})
\psi_h
	\;+\; \chi_h^\dagger ({\bf D} \cdot g {\bf E} - g {\bf E} \cdot
{\bf D})\chi_h\right) \nonumber \\
&+& \frac{1}{8m^2}
\left( \psi_h^\dagger (i {\bf D} \times g {\bf E} - g {\bf E} \times i {\bf D})
\cdot \mbox{ $\bsigma$} \psi_h
	\;+\; \chi_h^\dagger (i {\bf D} \times g {\bf E} - g {\bf E} \times
i {\bf D})
	\cdot \mbox{ $\bsigma$} \chi_h \right) \nonumber \\
&+& \frac{1}{2m}
\left( \psi_h^\dagger (g {\bf B} \cdot \mbox{ $\bsigma$}) \psi_h
	\;-\; \chi_h^\dagger (g {\bf B} \cdot \mbox{ $\bsigma$})
\chi_h
\right)+O(g^3)\,,
\end{eqnarray}
since covariant derivatives and $\bf E$ both consist of two terms of differing
orders in $v$.

\begin{figure}[tb]
\epsfxsize=10cm
\hfil\epsfbox{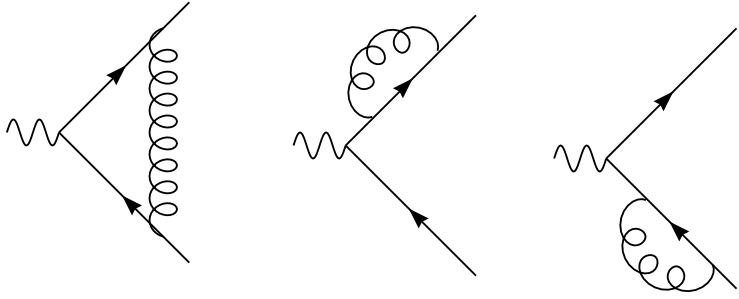}\hfill
\caption{One loop contributions to quark-antiquark production in QCD.}
\label{qcdloops}
\end{figure}

Since radiation gluons do not contribute to one-loop graphs, as was discussed
in the previous section, their couplings are not presented here.

Using the  relations (\ref{gammamatch}) and
\begin{equation}
\bar u(p_1)v(p_2)=-{1\over 2m}u_h^\dagger\; {\bf p}\cdot\bsigma v_h+O({\bf
p}^3)\,,
\end{equation}
(still working in the centre of mass frame), the amplitude for quark-antiquark
production in QCD from the diagrams in Fig. \ref{qcdloops} may be
expanded in powers of ${\bf p}/m$:
\begin{eqnarray}\label{qcdamp}
i{\cal A}_{\rm{QCD}} &=&u^\dagger_h\bsigma^i v_h\left(1-{2 g^2\over
3\pi^2}\right)-{1\over 2m^2} u^\dagger_h \pvec\cdot\bsigma \pvec^i
v_h\left(1-{g^2\over 3\pi^2}\right)\nonumber\\
&+&{g^2\over 12\pi^2}u^\dagger_h\bsigma^i v_h\left[
{m\over |\pvec|}\left( \pi^2 + i \pi\left( \gamma_E + {2\over d-4}  +
\ln{\pvec^2\over \pi\mu^2 } \right)\right)
\right. \nonumber\\
%& +&\left. \left( -3 \gamma_E-4 -{6\over d-4} -3 \ln{m^2\over 4\pi\mu^2}
%\right)\right. \nonumber\\
&&\qquad\qquad+ \left. {3|\pvec|\over 2m}\left(\pi^2+ i
\pi\left(\gamma_E-2 + {2\over d-4}  + \ln{\pvec^2\over \pi\mu^2}\right)\right)
\right. \nonumber\\
&&\qquad\qquad+ \left. {\pvec^2\over 3m^2} \left( {2\over 3} -8 \gamma_E -
{16\over d-4}  - 8\ln{m^2\over 4\pi\mu^2}\right)\right]\nonumber\\
&+&{g^2\over
12\pi^2}{u_h^\dagger\pvec\cdot\bsigma \pvec^i v_h\over 2m^2}\left[ {m\over
|\pvec|}\left(
-\pi^2 - i \pi\left( \gamma_E -2+ {2\over d-4}  +   \ln{\pvec^2\over \pi\mu^2 }
\right)\right)
\right]+O(v^3)\,,
\end{eqnarray}
(where an overall factor of $E/m$ has been divided out, to convert to
nonrelativistically normalized states).
The amplitude has the expected $1/v$ singularity from Coulomb scattering,
signalling the failure of perturbation theory close to threshold.

\begin{figure}[tb]
\epsfxsize=10 cm
\hfil\epsfbox{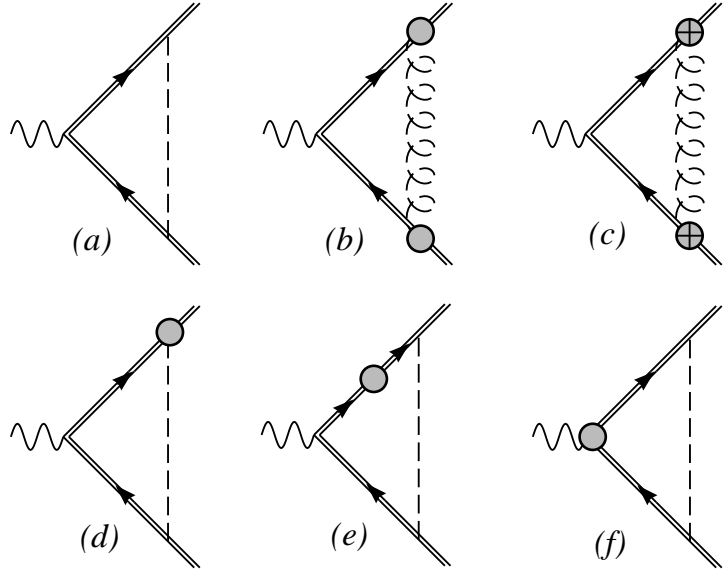}\hfill
\caption{One loop contributions to quark-antiquark production in
NRQCD.  The dashed line corresponds to a
potential $A_0$ gluon,
the dashed gluon line to a potential $A_i$ gluon.  The shaded circles represent
(b) the $\pvec\cdot {\bf A}$ vertex, (c) the Fermi vertex, (d) the Darwin
vertex, (e) the relativistic kinematic correction to the fermion leg, and
(f) ${\bf O}_2$.   Implicit in both (d) and (e) are graphs with
the same operator insertion on the antiquark line.  The wavefunction graphs
vanish.}
\label{nrqcdgraphs}
\end{figure}

The matching conditions for the current are given by the difference between
Eq.\ (\ref{qcdamp}) and the graphs in Fig. \ref{nrqcdgraphs} computed
on-shell in NRQCD \cite{onshell}.
The graph in Fig. \ref{nrqcdgraphs}(a) corresponds to Coulomb $A_P^0$ exchange.
The only subtlety in
evaluating this graph  is that, to the order in which we are working, the
on-shell condition
in the effective theory is
\begin{equation} T={\pvec^2\over 2m}-{\pvec^4\over 8m^3}\,,
\end{equation} instead of the leading order relation $T=\pvec^2/2m$, and this
extra term must be treated correctly as  a perturbation so as not to violate
power counting.  Evaluating the
graph in Fig. \ref{nrqcdgraphs}(a) for  arbitrary off-shell spinors
\begin{equation}
{2 mT \over \pvec^2}=1+\delta\,,
\end{equation}
gives
\begin{equation}
(a)=c_1{g^2 m^{d-6}\pvec^2\over 3 \cdot 2^{d-3}\pi^{{d-1}\over
2}}u^\dagger_h\bsigma^i v_h
\left(-{\pvec^2\over
m^2}+i\epsilon\right)^{{d-7}\over2}\delta^{{d-5}\over2}\Gamma\left({3-d\over
2}\right){}_2F_1\left({d-3\over2},{5-d\over 2},{d-1\over
2},-{1\over\delta}\right).
\end{equation}
The hypergeometric function ${}_2F_1(a,b,c;\xi)$ may be expanded in powers of
$\delta$, giving
\begin{eqnarray}
(a)&=&c_1{(d-3)g^2 m^{d-2}\over3(d-4)2^{d-2}\pvec^2\pi^{{d-1}\over
2}}u^\dagger_h\bsigma^i v_h \left(-{\pvec^2\over
m}+i\epsilon\right)^{{d-3}\over 2}\Gamma\left({3-d\over
2}\right)\nonumber \\ &&\times\left(1-\left(2-{d\over
2}\right)\delta+{(d-4)\Gamma(4-d)\Gamma\left({d-3\over 2}\right)\over
\Gamma\left({5-d\over 2}\right)}\delta^{d-4}+O(\delta^2)\right).
\end{eqnarray}
Evaluating this graph with the leading-order on-shell condition, $\delta=0$,
the $O(\delta^{d-4})$ term vanishes as long as $\mbox{Re}(d)>4$.  In
order to
have this result remain as the leading term in the expansion away from
$\delta=0$ so that
power counting is retained, the $\delta^{d-4}$ term must be evaluated in
dimensional
regularization as a formal power series, $f(\delta)=f(0)+\delta
f^\prime(0)+\dots$.   In
this case, each term in the expansion vanishes for sufficiently large
$\mbox{Re}(d)$, so the
entire series vanishes in dimensional regularization.  The final result for
this graph near
$d=4$ is therefore
\begin{equation}
(a)=c_1{g^2\over 12\pi^2}u^\dagger_h\bsigma^i v_h{m\over |\pvec|} \left[ \pi^2
+ i\pi\left(
\gamma_E + {2\over d-4}  +  \ln{\pvec^2\over \pi\mu^2
}+i\delta\right)+O(\delta^2) \right]\,,
\end{equation}
which, for $\delta=-\pvec^2/4m^2$, gives
\begin{equation} (a)=c_1{g^2\over 12\pi^2}u^\dagger_h\bsigma^i v_h\left[{m\over
|\pvec|} \left(
\pi^2 + i\pi\left( \gamma_E + {2\over d-4}  + \ln {\pvec^2\over \pi\mu^2
}\right)\right)-{i \pi\over 4}{|\pvec|\over m}\right].
\end{equation}
This reproduces the $O(1/v)$ term in the full amplitude.

As discussed above, there are
no graphs at $O(g^2 v^{2n})$ in NRQCD from radiation gluon loops. At $O(g^2 v)$
there are contributions from the leading
relativistic corrections to Coulomb scattering.  In addition, since Coulomb
exchange scales as $v^{-1}$, the dressing of ${\bf O}_2$ with a single
$A^0_P$ exchange also contributes at $O(g^2 v)$.
$A^i_P$ exchange contributes both via the
$\pvec\cdot{\bf A}$ coupling
(Fig. \ref{nrqcdgraphs}(b))
\begin{equation} (b)=c_1{g^2\over 12\pi^2}u^\dagger_h\bsigma^i v_h{|\pvec|\over
m} \left( \pi^2 +
i\pi\left( \gamma_E -1+ {2\over d-4}  + \ln {\pvec^2\over \pi\mu^2
}\right)\right).
\end{equation} and the Fermi coupling
\begin{equation}
(c)=c_1{g^2\over
12\pi^2}\left(u^\dagger_h\bsigma^i v_h{|\pvec|\over
m}+{m\over|\pvec|}{u^\dagger_h\pvec\cdot\bsigma \pvec^i v_h\over
m^2}\right)\left(-{i\pi\over 2}\right)\,.
\end{equation}
Coulomb exchange is corrected by the Darwin vertex,
\begin{equation}
(d)=c_1{g^2\over 12\pi^2}u^\dagger_h\bsigma^i v_h{|\pvec|\over
m}\left(-i\pi\right)\,,
\end{equation}
while the spin-orbit coupling does not contribute.  The relativistic
corrections to the quark and antiquark propagators give
\begin{equation}
(e)=c_1{g^2\over 12\pi^2}u^\dagger_h\bsigma^i v_h{|\pvec|\over
2m}\left[\pi^2+i\pi\left(\gamma_E+{1\over 2}+{2\over
d-4}+\ln{\pvec^2\over\pi\mu^2}\right)\right]\,,
\end{equation}
and finally, the one-loop correction to ${\bf O}_2$ in Fig.
\ref{nrqcdgraphs}(f) gives
\begin{eqnarray}
(f)&=&-c_2{g^2\over 12\pi^2}{u^\dagger_h\pvec\cdot\bsigma\pvec^i v_h\over
m^2}{m\over
2|\pvec|} \left( \pi^2 + i\pi\left( \gamma_E -3+ {2\over d-4}  + \ln
{\pvec^2\over \pi\mu^2
}\right)\right)\nonumber \\ &-&c_2{g^2\over
12\pi^2}u^\dagger_h\bsigma^i v_h{|\pvec|\over m}\left({i\pi\over 2}\right).
\end{eqnarray}
Combining these results gives
\begin{eqnarray}\label{nrqcdamp}
i{\cal A}_{\rm{NRQCD}}&=&c_1u^\dagger_h\bsigma^i v_h-{c_2\over
2m^2}u^\dagger_h\pvec\cdot\bsigma \pvec^i v_h\nonumber\\
&&+{g^2\over 12\pi^2} u^\dagger_h\bsigma^i v_h \left[ c_1{m\over|\pvec|}\left(
\pi^2
+ i \pi\left( \gamma_E + {2\over d-4}  +   \ln{\pvec^2\over \pi\mu^2 }
\right)\right)
\right. \nonumber\\
&&\qquad\qquad+ \left. {3|\pvec|\over 2m}\left(c_1\left(\pi^2+ i
\pi\left(\gamma_E-{5\over 3} +
{2\over d-4}  + \ln{\pvec^2\over \pi\mu^2}\right)\right)-{i\pi\over
3}c_2\right)
\right] \\
&&+{g^2\over 12\pi^2} {u^\dagger_h\pvec\cdot\bsigma \pvec^i v_h\over
2m^2}\left[
{m\over|\pvec|}\left(-c_2\left(\pi^2+ i \pi\left(\gamma_E-3 +
{2\over d-4}  + \ln{\pvec^2\over \pi\mu^2}\right)\right)-i\pi
c_1\right)\right].\nonumber
\end{eqnarray}
As required, all the nonanalytic dependence on the external momentum cancels in
the matching.  This result is also gauge independent.

Comparing Eqs.\ (\ref{qcdamp}) and (\ref{nrqcdamp}) gives the
coefficients $c_1 - c_3$ (regulating the low-energy theory as usual in
$\overline{\mbox{MS}}$)
to $O(g^2)$:
\begin{eqnarray}
c_1&=&1-{8\alpha_s\over 3\pi}+O(\alpha_s^2)\nonumber\\
c_2&=&1-{4\alpha_s\over 3\pi}+O(\alpha_s^2)\nonumber\\
c_3&=&-{\alpha_s\over 9\pi}\left( {2\over 3} - 8\ln{m^2\over \mu^2}\right)
+O(\alpha_s^2).
\end{eqnarray}
The result for $c_1$ reproduces the familiar short-distance correction to
$e^+ e^-\rightarrow q\bar q$ near threshold \cite{sdist},
whereas $c_2$ and $c_3$ generalize this to $O(v^2)$.  Note that
the bare $c_1$ is finite while the bare $c_3$ is divergent.  This reflects
the fact that there
are no infrared or ultraviolet divergences in the amplitude
at $O(v^0)$ since the quarks are in a colour singlet state, and therefore
cannot radiate a gluon
at leading order in the multipole expansion.

The major result of this section is that the nonanalytic dependence on the
external momenta in the QCD amplitude is exactly reproduced in NRQCD.
This provides a
nontrivial check of the consistency of this approach beyond leading order.
However, only for values of the coupling and external momenta such that
$\alpha_s\ll v\ll 1$ does the tree-level matching of ${\bf O}_2$ and ${\bf
O}_3$
dominate the two-loop matching of ${\bf O}_1$.
For scattering states closer to threshold (as well as for bound states) where
$v\lesssim \alpha_s$, ladder graphs containing potential gluons
must be summed to all orders via the Schr\" odinger equation.
In this case,
graphs containing a single insertion of the tree-level matching of ${\bf O}_2$,
the two-loop matching of ${\bf O}_2$ and  the tree-level matching of  ${\bf
O}_1$ combined
with a single higher order potential contribution
are equally important.
In this region, the one-loop matching to ${\bf O}_2$ and ${\bf O}_3$ that
are presented here are the same  order as the three  loop matching to ${\bf
O}_1$.

\section{Conclusions}\label{conc}

In this paper we have presented a power counting scheme for nonrelativistic
effective theories that
allows for a systematic calculation of subleading effects.  A systematic $v$
counting
scheme simplifies
the calculation of relativistic corrections to QCD processes such as
quarkonium production and decay.   As it is usually presented, NRQCD does not
have manifest
$v$ power counting in
the Lagrangian, nor is $v$ power counting preserved by loop graphs either in
dimensional
regularization or with a momentum cutoff.  While there is nothing in principle
wrong with
this, it makes calculating matching conditions somewhat awkward, since the
matching
conditions for any given operator will change by $O(1)$ when higher order
operators
are included in the Lagrangian.

Velocity power counting is only preserved by loop
graphs in dimensionally regulated NRQCD when gluons contributing to
potential scattering are treated separately from
on-shell gluons.  This was accomplished in this paper by introducing two
distinct gluon
fields in the effective theory.
Potential gluons propagate instantaneously and give rise to the QCD potential,
whereas radiation gluons do
not contribute to potential scattering, but correspond to on-shell gluons.
The power counting is manifest in the Lagrangian when space, time and
the fields are rescaled
for the potential fields as discussed in Ref.\ \cite{lm96} and for
radiation fields as discussed
in Ref.\ \cite{gr97}.  As shown in Ref.\ \cite{gr97}, under this rescaling
radiation fields couple to fermions
via the multipole expansion.  Separating these gluon modes realizes at the
level of the Lagrangian the separation advocated for NRQED in Ref.\
\cite{la96}.
Under this rescaling $v$ power counting is manifest in any gauge, not just
Coulomb gauge, and also holds for non-gauge interactions.

The infrared divergences arising in fermion-antifermion
production in Yukawa theory at order $v^{-1}$ and $v^0$ were shown to be
reproduced in the
nonrelativistic effective theory only when both potential and
radiation scalars were included, and the matching conditions at that order were
shown to be analytic in the external
momentum.  Finally, the matching conditions for quark-antiquark production by
an external vector current were computed in NRQCD to $O(g^2 v^2)$.

\section*{Acknowledgments}

We are particularly grateful to Aneesh Manohar for many discussions on the
subject of this work.  We also thank Adam Falk, David Kaplan, Ira Rothstein
and Larry Yaffe for useful discussions.  This work was supported
in part by the Natural Sciences and Engineering Research Council of Canada
and by United States Department of Energy under Grant No.\ DOE/ER/41014-21-N97.
M.\ L.\ acknowledges additional support by the Alfred P. Sloan Foundation.
M.\ L.\ thanks the Nuclear Theory Group of the University of Washington for
their hospitality during his stay.


\begin{references}

\bibitem{nrqcd}
W.E. Caswell and G.P. Lepage, Phys. Lett. {\bf B167}, 437 (1986);
G.T. Bodwin, E. Braaten, and G.P. Lepage, Phys. Rev. {\bf D51}, 1125 (1995);
Phys. Rev. {\bf
D55}, 5853 (1997) (E).

\bibitem{hqet} N. Isgur and M.B. Wise, Phys. Lett. {\bf B232}, 113 (1989);
Phys.
Lett. {\bf B237}
527 (1990); H. Georgi,  Phys. Lett. {\bf B240},  447 (1990);
E. Eichten and B. Hill, Phys. Lett. {\bf B243},  427 (1990).

\bibitem{nrsum}  V. A. Novikov et.\ al., Phys. Rept. {\bf 41}, 1 (1978);
M. B. Voloshin,
Int. J. Mod. Phys. {\bf A10}, 2865 (1995).

\bibitem{posit} W. E. Caswell and G. P. Lepage, Phys. Rev. {\bf A20}, 36
(1979); P. Labelle,
G. P. Lepage and U. Magnea, Phys. Rev. Lett. {\bf 72}, 2006 (1994).

\bibitem{nnscatt} S. Weinberg, Phys. Lett. {\bf B51}, 288 (1990); Nucl. Phys.
{\bf B363}, 3 (1991);
Phys. Lett. {\bf B295},114 (1992); C. Ordonez, U. van Kolck, Phys. Lett.
{\bf B291},
(1992) 459; C. Ordonez,
L. Ray,  U. van Kolck, Phys. Rev. Lett. {\bf 72}, (1994) 1982; Phys. Rev.
{\bf C53}, (1996)
2086.;  U. van Kolck, Phys.
Rev. {\bf C49},  (1994) 2932; D.B. Kaplan, M.J. Savage and M.B. Wise,
Nucl. Phys.
{\bf B478}, 629 (1996).

\bibitem{hqetmatch} A. Falk, H. Georgi, B. Grinstein and M. B. Wise, Nucl.
Phys.
{\bf B343}, 1 (1990);
A. Falk and B. Grinstein, Phys. Lett. {\bf B249}, 314 (1990);
M. Neubert, Nucl. Phys. {\bf B371}, 149 (1992);
A. F. Falk and M. Neubert,  Phys. Rev. {\bf D47}, 2965 (1993).

\bibitem{hqetcomplex} B. Grinstein, W. Kilian, T. Mannel and M. B. Wise,
Nucl. Phys. {\bf B363} 19 (1991).

\bibitem{le92} G.P. Lepage et.\ al., Phys. Rev. {\bf D46}, 4052 (1992).

\bibitem{la96} P. Labelle, McGill Preprint McGill/96-33 (1996), hep-ph/9608491.

\bibitem{lm96} M. Luke and A. V. Manohar, Phys. Rev. {\bf D55}, 4129 (1997).

\bibitem{gr97} B. Grinstein and I. Z. Rothstein, hep-ph/9703298 (1997).

\bibitem{sj92} A. Falk, H. Georgi, B. Grinstein and M.B. Wise, Nucl. Phys. {\bf
B343}, 1 (1990),  E. Jenkins and M.J. Savage, Phys. Lett. {\bf B281}, 331
(1992); J.
Goity,
Phys. Rev. {\bf D46}, 3929 (1992).

\bibitem{ah97} A. Hoang, hep-ph/9702331  (1997);  hep-ph/9704325  (1997).

\bibitem{am97} A. V. Manohar, Phys. Rev. {\bf D56}, 230 (1997).

\bibitem{onshell} H. Georgi, Nucl. Phys. {\bf B361}, 339  (1991).

\bibitem{sdist} R. Karplus and A. Klein, Phys. Rev. {\bf 87}, 848 (1952);
R. Barbieri,
R. Gatto, R. K\" ogerler and Z. Kunzst, Phys. Lett. {\bf B57},455  (1975).
\end{references}
\end{document}